%% file: main.tex
\documentclass[conference]{IEEEtran}

\pdfoutput=1 

\ifCLASSINFOpdf
  \usepackage[pdftex]{graphicx}
  \graphicspath{{../}{./}}

  \makeatletter
  \def\input@path{{./}{conference/}}
  \makeatother

\else
\fi
\usepackage{url}

\usepackage{relsize}
\usepackage{xspace}

\usepackage{booktabs} 

\usepackage{caption}
\usepackage{subfig}
\usepackage{multirow}

\usepackage{todonotes}

\usepackage{amsthm}
\usepackage{amsmath}
\usepackage{amssymb}
\usepackage{etoolbox}
\usepackage{cuted}

\usepackage{tabularx}

\theoremstyle{definition}

\newtheorem{proposition}{Proposition}[section]
\newtheorem{theorem}{Theorem}[section]

\newtheorem{corollary}[theorem]{Corollary}

\usepackage[ruled]{algorithm2e}
\usepackage{enumitem}

\input{macros}

\newcommand{\Conv}{%
  \mathop{\scalebox{1.0}{\raisebox{0.0ex}{$\circledast$}}
  }
 }

\begin{document}
\title{Towards a Performance Model for Byzantine Fault Tolerant (Storage) Services}

\author{
  \IEEEauthorblockN{Thomas Loruenser, Benjamin Rainer, Florian Wohner}
  \IEEEauthorblockA{
  Center for Digital Safety \& Security\\
  Austrian Institute of Technology (AIT)\\
  Vienna, Austria\\
  Email: FirstName.LastName@ait.ac.at}
}

\maketitle


\begin{abstract}
Byzantine fault-tolerant systems have been researched for more than four decades, and although shown possible early, the solutions were impractical for a long time.
With PBFT the first practical solution was proposed in 1999 and spawned new research which culminated in novel applications using it today.
Although the safety and liveness properties of PBFT-type protocols have been rigorously analyzed, when it comes to practical performance only empirical results - often in artificial settings - are known and imperfections on the communication channels are not specifically considered.
In this work we present the first performance model for PBFT specifically considering the impact of unreliable channels and the use of different transport protocols over them.
We also did extensive simulations to verify the model and to gain more insight on the impact of deployment parameters on the overall transaction time.
We show that the usage of UDP can lead to significant speedup for PBFT protocols compared to TCP when tuned accordingly even over lossy channels.
Finally, we compared the simulation to a real implementation and measure the benefits of a developed improvement directly.
We found that the impact on the design of the network layer has been overlooked in the past but offers some additional room for improvement when it comes to practical performance.
In this work we are focusing on the optimistic case with no node failures, as this is hopefully the most relevant situation.

\end{abstract}


\maketitle

\input{sections/introduction}
\input{sections/motivation}

\input{sections/related_work}
\input{sections/model}

\input{sections/eval}
\input{sections/conclusio}


\bibliographystyle{abbrv}
\bibliography{literature,literature-lot,literature-soft}

\end{document}

%% file: conference/macros.tex
\newcounter{SSschemeCounter}                         

\let\oldSSschemeCounter\theSSschemeCounter
\renewcommand{\theSSschemeCounter}{\thesection.\oldSSschemeCounter}
{  \end{description}%
   \hrule height .01cm
   \end{revision}\medskip}

\usepackage{cleveref}
\crefformat{equation}{(#2#1#3)}
\Crefformat{equation}{Equation~(#2#1#3)}
\crefrangeformat{equation}{(#3#1#4) to~(#5#2#6)}
\crefmultiformat{equation}{(#2#1#3)}{ and~(#2#1#3)}{, (#2#1#3)}{ and~(#2#1#3)}
\crefrangemultiformat{equation}{(#3#1#4) to~(#5#2#6)}{ and~(#3#1#4) to~(#5#2#6)}{, (#3#1#4) to~(#5#2#6)}{ and~(#3#1#4) to~(#5#2#6)}
\crefformat{enumi}{(#2#1#3)}
\Crefformat{enumi}{(#2#1#3)}
\Crefrangeformat{equation}{Equations~(#3#1#4) to~(#5#2#6)}
\Crefmultiformat{equation}{Equations~(#2#1#3)}{ and~(#2#1#3)}{, (#2#1#3)}{ and~(#2#1#3)}
\Crefrangemultiformat{equation}{Equations~(#3#1#4) to~(#5#2#6)}{ and~(#3#1#4) to~(#5#2#6)}{, (#3#1#4) to~(#5#2#6)}{ and~(#3#1#4) to~(#5#2#6)}
\crefname{example}{Example}{Examples}
\Crefname{example}{Example}{Examples}
\crefname{figure}{Figure}{Figures}
\Crefname{figure}{Figure}{Figures}
\crefname{definition}{Definition}{Definitions}
\Crefname{definition}{Definition}{Definitions}
\crefname{SSschemeCounter}{Algorithm}{Algorithms}
\Crefname{SSschemeCounter}{Algorithm}{Algorithms}

%% file: conference/sections/introduction.tex
\section{Introduction}
Cloud services have become pervasive in our daily life.
We use cloud services to store and share data (e.g., photos, videos) with friends, family, and colleagues.
Companies rely on cloud storage services because they provide a reasonable and convenient (from a monetary/maintaining point of view) alternative to in-house storage solutions.
Although the availability and durability of individual offerings can be quite good,
combining them into virtual multi-cloud storage could be very appealing but challenging, especially if the connectivity is not ideal and high robustness is needed.

Typically, protocols that tolerate Byzantine faults are needed in this setting.
However, implementing well-performing solutions has proven challenging.
The most promising approaches are based on Practical Byzantine Fault Tolerance (PBFT), originally introduced by \cite{Castro:1999:PBF}.
PBFT is a 3-phase protocol that relies only on a weak synchrony assumption to guarantee safety and liveness even over unreliable channels.
It is known to perform well in local LAN settings with high-bandwidth connectivity and low latency, but we found the achieved performance in typical multi-cloud settings disappointing.

In this work we take a deep dive into the network layer and protocols for PBFT implementations for lossy and medium to high latency channels.
To the best of our knowledge, we present the first approach for a performance model of PBFT.
We analyze the core 3-phase view-consensus protocol in PBFT without additional features like leader change and checkpointing and develop an analytical performance model for success probability and latency of transactions.
Then we present simulation results and analyze systems performance using TCP and UDP as transport protocols.
We further explore the parameters available for tuning such systems and evaluate the model with extensive simulations and provide criteria for system design and a hybrid transport mode that is able to increase performance by making use of both TCP and UDP.
The results are then compared to a real implementation in a comparable environment.

The remainder of the paper is organized as follows.
In the rest of this section we briefly discuss our motivation and relevant related work.
In Section~\ref{sec:model} we present the analytical model and Section~\ref{sec:performance_evaluation} provides a performance evaluation of our service in different configurations.
Section~\ref{sec:con} summarizes the paper and provides an outlook on future work.

%% file: conference/sections/motivation.tex
\subsection{Motivation}
\label{sec:motivation}

Our analysis was motivated by the performance problems encountered in the deployment and operation of robust and secure multi-cloud storage solutions in the spirit of \cite{Loruenser:2015:ARCHISTAR,Padilha2011}, which suffer from the worse connectivity compared to LAN settings.
A multi-cloud deployment over different administrative domains (clouds) has the advantage over pure LAN deployments that there is no need to fully trust a single provider.
Nevertheless, the setting is not fully untrusted like Blockchain applications, where nodes do not have any trust relations at all.
In that sense our application scenario is somewhere in between fully-trusted and fully-untrusted, which we feel is a very reasonable assumption to build trustworthy data spaces, e.g. for governmental or healthcare solutions.

In our scenario we have to cope with less reliable networks than pure LAN implementations, but still reasonable connectivity, especially in the optimistic cases without node failures.
PBFT is designed for exactly this type of channel with weak synchrony, where messages are eventually delivered after a certain time bound $\Delta t$ which can be time-varying but is known to the protocol designer.
Moreover, it can also cope with unreliable channels.
Even better, the safety properties of PBFT hold even when the delay is violated, and only its liveness guarantees depend on the weak synchrony assumptions.
In fact, PBFT is a leader based consensus protocol with a 3-phase epoch (or view) consensus for safety in asynchronous networks and weak leader election mechanism to achieve progress, i.e., it is a good compromise for our use case.
Nevertheless, even when the weak synchrony assumptions hold, weakly synchronous protocols degrade significantly in throughput when the underlying network is unpredictable or unreliable.
Ideally, we would like a protocol whose throughput closely tracks the network's performance especially for the optimal case of no faults, but under the assumption of unreliable transport.

We encountered some of the above-described behavior in a concrete practical multi-cloud storage we built with PBFT where we compared the performance of the same implementation, once in a LAN-based setup and once in a cloud-based setup with Amazon Web Services (AWS).
The experiments clearly showed upload and download latencies for the AWS case to be worse than could be expected.
We suspected the reason for this discrepancy to be somewhere in the interplay of TCP, TLS, congestions, and higher latency.
Although this was only within a single cloud, the transaction times we experienced in our system were already unacceptable, and we decided to have a closer look at the dependencies of transaction times on network parameters like packet loss and latency, as multi-cloud settings will experience even rougher networking conditions.

%



%% file: conference/sections/related_work.tex
\subsection{Related Work}
\label{sec:rw}


When analyzing cloud data storage, two classes of failures are prominent: Byzantine- and crash-faults.
The latter describe systems that either work correctly or do not respond at all after an (initial) failure.
In contrast, Byzantine faults allow for arbitrary failures and thus do not limit an attacker's capabilities.
This makes them well suited for our approach which models cloud storage of sensitive data as subject to arbitrary faults, e.g., malicious attackers, network outages or memory corruption.

A commonly used protocol in the Byzantine setting is Practical BFT (PBFT)~\cite{Castro:1999:PBF} and its variants Zyzzyva \cite{zyzzyva} and Aardvark \cite{Clement2009a}.
It is leader based and utilizes majority voting between all involved servers and strong cryptography to provide message ordering and strong consistency in the face of Byzantine faults.
To allow for majority voting, active servers with communication channels between them are mandatory.
If privacy is additionally needed, the extensions of~\cite{Loruenser:2015:ARCHISTAR} show how it can be integrated with secret sharing.

Two types of deployments or applications can typically be distinguished, LAN and Blockchain.
If deployed in a closed network within a single administrative domain, i.e. in a LAN like for example the ''5 Chubby nodes within Google'' environment, best performance is achieved with the usage of UDP for message transmission.
To guarantee liveness in the face of errors and transactions not making progress, every PBFT transaction is associated to a ``view'' with a dedicated primary.
Whenever a transaction fails to make progress or a leader is suspected by the nodes of being malicious, a view-change is initiated, to prevent possibly faulty primaries from stalling transaction processing.
In a view change a new leader is elected and operation continues with this new leader, and for consistency reasons it is guaranteed that already committed transactions in the old view are correctly handled by nodes in the new view.
However, as the experiments of~\cite{chondros:hal-01555557} showed, due to congestion, packet loss can occur even in the ideal LAN-setting, and the triggered view-changes severely degrade performance.

In the blockchain world, different assumptions and requirements hold \cite{Kwon2014}, \cite{Yin2018} and \cite{Miller2016}, and results cannot easily be ported from one world to the other.
Many transactions are typically batched, and consensus is organized in epochs comprising all currently pending transactions.
Moreover, transaction times are typically amortized values, which makes sense in the blockchain setting with a continuous incoming stream of transactions and enough buffered transaction in each epoch.
The models also assume that a reliable channel can always be established with little overhead over unreliable channels and that the network buffers at nodes are infinite.
In practice, they typically apply TCP or its secure variant TLS if authenticity is required.

When it comes to performance analysis of BFT protocols, benchmarking is typically used to compare and estimate the performance of protocols \cite{Gupta2016}.
The only known more systematic approach was presented in \cite{Sukhwani2017}, which use Stochastic Reward Nets (SRN) to model “mean time to complete consensus”.
However, they model the network as a reliable channel where the rate of message transmission between all pairs of peers is the same and fit individual distributions from measurements.

In summary, a large body of research exists in BFT and many protocols have been proposed and benchmarked, but only little is known when it comes to performance modeling of such protocols, specifically considering the impact of the underlying communication channels, i.e., the impact of physical channel characteristics in the form of packet loss or latency.
Additionally, only few results are accompanied with open software implementations, which makes verification and comparison of results often hard or impossible.

%% file: conference/sections/model.tex
\section{Modeling the Impact of Packet Loss}
\label{sec:model}

In this section we briefly review the PBFT protocol and develop a performance model that specifically considers unreliable communication channels, which is always the case in real systems.
Due to space constraints we will focus on modeling transaction success and leave the model of transaction latency for the extended version.
We therefore compare the usage of the UDP and TCP transport protocols, and their impact on the performance of basic PBFT transactions.
For our analysis we look at the optimistic case, when no nodes are malicious, which is hopefully the most important case in a system.
However, the results can easily be extended to the full Byzantine case, as we will argue below.

\subsection{PBFT protocol}

\begin{figure}[tbh!]
\centering
\includegraphics[width=0.4\textwidth]{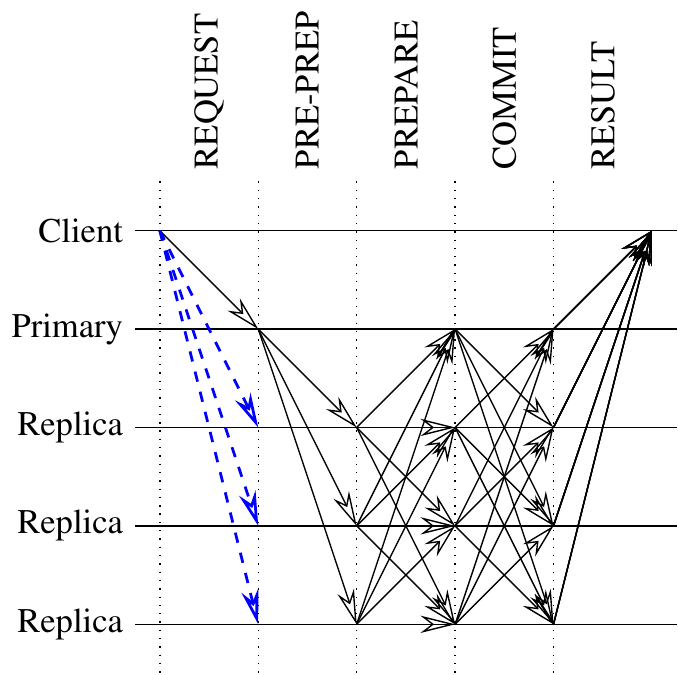}
\caption{Message flow diagram of PBFT with the extended first phase. Altered phases and communications are highlighted by red.}
\label{fig:msg-flow}
\end{figure}

PBFT basically resembles a state replication mechanism that can work over unreliable channels and guarantee safety and liveness even in asynchronous environments such as the Internet.
For this, it needs at minimum $3f+1$ nodes, tolerating up to $f$ of them being arbitrarily faulty in the Byzantine model.
The full protocol is leader-based as shown in Figure \ref{fig:msg-flow}, and the core view-consensus protocol comprises three phases which on a high level work as follows.
Being leader-based, one node takes over leadership in linearizing transactions for a given period of time, the so-called view, which can also be changed if enough nodes are not satisfied with the current leader (view-change).
During a view, the leader is getting transaction requests from clients and orders them by assigning a transaction identifier.
However, because the client does not know the current leader, it sends the request to all nodes.
For our analysis, which is only looking at the performance of the leader consensus, this part shown in blue can be omitted.
Having received the request, the leader broadcasts a PRE-PREPARE message.
If nodes receive a PRE-PREPARE they check transaction data and send a PREPARE message to all other nodes if it is consistent with their state.
If nodes receive enough PREPARE messages from other nodes they enter the prepared state and send a COMMIT message to all other nodes.
A node transitions into the committed state if it has received at least $2f+1$ (also including its own) COMMIT messages, and finally send a REPLY message to the client.
The client considers the transaction to be committed when it has received $f+1$ REPLY messages.

The protocol provides safety by only progressing if an honest majority is assured (at least $2f+1$ nodes are in the same state).
Furthermore, the commit phase is used to guarantee this property within views and the commit phase is needed to assure it over view changes.
Finally, liveness is guaranteed if the network satisfies weak synchrony conditions, which is often a reasonable assumption but could lead to large timeouts in software implementations and bad performance when the right timeout has to be found.
Weak synchrony means that eventually after a bounded time $\Delta t$ the network becomes synchronous, i.e., it eventually makes progress.

The PBFT protocol also applies cryptography to implement authentic channels and transaction certificates.
This protects from a malicious adversary with access to communication channels, even between honest nodes.
From a practical perspective, however, an attacker will not always have control over all communication channels between all nodes and this model seems unnecessarily restrictive when it comes to performance evaluation.
For real-world applications, especially in the multi-cloud setting, it is therefore reasonable to assume that if an attacker compromises one node, it has full control over it and can control all incoming and outgoing messages of that node, but it is not able to control the channels between honest nodes.
In such a model, the are some redundancies in the PBFT protocol that can be safely optimized to achieve best performance for imperfect channels.
Contrarywise, in a model where the adversary also transports all messages, such a treatment is less helpful.
If the adversary can arbitrarily delay all messages, network optimization is not possible.

In our work, as a first step we will model and analyze the optimistic case with no malicious nodes present but possibly adverse and unreliable network conditions as we assume this to be the typical case.
The goal of this first approach is to fully leverage the redundancy inherent in PBFT to achieve short transaction times in optimal cases.
Note that in the case of errors we can always fall back to a standard implementation for non-optimistic case with known performance degradation.

\subsection{Modeling Transaction Success}

As mentioned before, if requests time-out a view change is triggered. These view changes inflict high resource costs (especially on the network level); in addition new requests can only be executed after the view change has been completed. Thus, it would be beneficial to know (or at least estimate) the probability that the system is able to successfully process a request a priori. This knowledge could significantly improve the overall system performance because if an unreliable transport mechanism, i.e., UDP, is used the system may switch over to reliable network communication, i.e., TCP, if the chance of a view change increases.

The employed PBFT protocol heavily relies on network communication between the replicas. Thus, delay and packet loss can have a tremendous impact on the overall system performance. There are basically two transport protocols: UDP (connectionless) and TCP (connection orientated).
Both protocols are suited for our system (both provide disadvantages and advantages), however, UDP employs the least overhead and delay while TCP requires maintaining a connection and provides a reliable transport service. In order to minimize communication overhead and delay, UDP is favored.
However, with increasing packet loss, we may run into the problem that nodes do not receive at least $2f+1$ messages from other nodes in a phase (cf. Figure~\ref{fig:msg-flow}). If this applies to more than $2f+1$ nodes, phases cannot be accepted anymore because of missing (distinct) messages and, therefore, requests will time-out.
This leads to re-requesting timed-out requests and finally ends in even more requests timing-out. Thus, if the packet loss increases, TCP intuitively becomes superior to UDP, while trading performance for reliability. Thus, the question ``when should TCP be used instead of UDP?'' arises. For the following considerations $f \in [0, \lfloor\frac{n-1}{3}\rfloor]$, in order to have more than $f$ correct working replicas we need $n-2f>f \Rightarrow n>3f$ replicas, thus the smallest number of needed replicas is $3f+1$ assuming $f$ faulty ones.
In the following we will provide a criteria which answers the aforementioned question based on probability theory.

Intuition tells us, that we would switch over to TCP if the expected number of nodes that receives more than $2f+1$ message is less than $2f+1$ in order to have enough replicas transitioning between the declared PBFT phases. Our goal is it to investigate how errors in the actual transmission between the BFT protocol phases propagate and how these errors influence the successful completion of a given transaction under the assumption of $f$ faulty nodes. Without loss of generality, we assume that multicast is not in place and, therefore, nodes have to rely on broadcasts. If messages are attacked by man-in-the-middle attacks and are altered (thus altering the recalculated digest) we assume that the message is lost.

Taking a look at Figure~\ref{fig:msg-flow} and having in mind that messages may get lost we have the following phases if a request is received by the primary:
\begin{enumerate}[label=(\roman*)]
\item \textbf{PRE-PREPARE}: The primary sends a PRE-PREPARE message to all nodes (including itself). Nodes can only successfully commit a transaction if they successfully accept all phases this also includes the reception of a PRE-PREPARE message which actually fires off the consensus protocol. Assuming that there is packet loss, $m$ out of $n-1$ ($m, n \in \mathbb{N}, m \leq n$) nodes may receive a PRE-PREPARE message. The primary itself sends $n-1$ PRE-PREPARE messages to only $n-1$ nodes.
\item \textbf{PREPARE}: $m+1$ (accounting for the primary) nodes broadcast a PREPARE message to all $n$ nodes. Each node has to receive at leas $2f+1$ PREPARE messages to successfully accept the PREPARE phase and in order to transition into the next phase. We start with $m+1$ nodes and may end up with only $k$ out of $m+1$ nodes ($k,m, n \in \mathbb{N}, k \leq m \leq n$) receiving at least $2f+1$ PREPARE messages. A node in this phase will only need to receive $2f$ distinct PREPARE messages from $m$ nodes because one message is send to itself.
\item \textbf{COMMIT}: $k$ nodes transition into this phase and broadcast a COMMIT message to all $n$ nodes. Since only $k$ nodes successfully accepted the previous phase we again have at most $k$ nodes which can successfully accept the last phase. Thus, we have $j$ out of $k$ nodes ($j,k,m, n \in \mathbb{N}, j \leq k \leq m \leq n$) which again need $2f$ messages from $k-1$ nodes.
\item \textbf{REPLY}: $j$ nodes arrive in this phase and will send a REPLY to the client. The client sees its request as fulfilled if it receives $f+1$ (best case) or $2f+1$ (worst case) REPLY messages out of $j$ possible ones.
\end{enumerate}
We denote the random variables for the phases as follows: $M$ (PRE-PREPARE), $K$ (PREPARE), $J$ (COMMIT), and $S$ (REPLY). We do not take into account the reception of a request. If a request is not received, no transaction will be triggered. The final number of nodes, thus, relies on the number of nodes that are able to successfully accept each phase. We assume that the probability of successfully transmitting a packet is independent and identically distributed.
\begin{figure*}
\begin{equation}
\label{eq:probability_nodes}
\begin{split}
\mathbb{P}(S=s, J=j ,K=k, M=m)=  \left(
    \begin{array}{c}
      j \\
      s
    \end{array}
  \right) p_l^s (1-p_l)^{j-s} \left(
   \begin{array}{c}
      k \\
      j
    \end{array}
  \right) \mathbb{P}_{T}(X \geq 2f| k-1)^j (1-\mathbb{P}_{T}(X \geq 2f|k-1))^{k-j} \\
  \left( \begin{array}{c}
      m+1 \\
      k
    \end{array} \right) \mathbb{P}_{T}( X \geq 2f| m)^k (1-\mathbb{P}_{T}(X \geq 2f| m))^{m+1-k} \left(\begin{array}{c}
      n-1\\
      m
    \end{array}
    \right) p_l^m (1-p_l)^{n-1-m} 
    \end{split}
\end{equation}
\end{figure*}
The expected value $\mathbb{E}[S, J \geq 2f+1, K \geq 2f+1, M \geq 2f+1]$ as a function of successful transmitting a message/packet, should suffice the following properties:
\begin{enumerate}[label=(\roman*)]
\item Let $f \in [0, \lfloor\frac{n-1}{3}\rfloor]$, $\forall p_{l,i}, p_{l,j} \in \quad ]0,1[, p_{l,i} \leq p_{l,j}: \mathbb{E}[S, J \geq 2f+1, K \geq 2f+1, M \geq 2f+1](p_{l,i}) \leq \mathbb{E}[S, J \geq 2f+1, K \geq 2f+1, M \geq 2f+1](p_{l,j})$.
\item Let $p_l \in \quad ]0,1[: \mathbb{E}[S, J \geq 2f_i+1, K \geq 2f_i+1, M \geq 2f_i+1] \geq \mathbb{E}[S, J \geq 2f_j+1, K \geq 2f_j+1, M \geq 2f_j+1]$, $\forall f_i, f_j \in [0, \lfloor\frac{n-1}{3}\rfloor], f_i \leq f_j$.
\end{enumerate}
The probability that a client receives $s$ replies from $j$ nodes, where $l$ out of $k$ nodes accepted the COMMIT phase, $k$ out of $m$ nodes accepted the PREPARE phase, and $m$ out of $n$ nodes successfully received a PRE-PREPARE message is given by Equation~\ref{eq:probability_nodes}. We define $p_l$ as the probability for successfully transmitting a packet with length $l$ (we will later derive this probability or provide means to measure it). The actual probability $p_l$ does depend on the underlying transport protocol $T$. Furthermore, $\mathbb{P}_T(X=k|n,p_l)$ denotes the probability that $k$ out of $n$ packets/messages are successfully transmitted given $p_l$ using transport protocol $T$. The following result provides an estimate of the expect value which can be evaluated fast.
\begin{proposition}
\label{proposition:expected_value}
Let $1 \leq f \leq \lfloor \frac{n-1}{3} \rfloor$, then we have the following inequality for $\mathbb{E}[S, J \geq 2f+1, K \geq 2f+1, M \geq 2f+1]$:
%
\begin{equation}
\label{eq:expected-value}
\begin{split}
&\mathbb{E}[S, J \geq 2f+1, K \geq 2f+1, M \geq 2f+1] \geq p_l^2 n \\
& \sum_{m=0}^{n-2}
    \left(\begin{array}{c}
      n-2\\
      m
    \end{array}
    \right) p_l^{m} (1-p_l)^{n-2-m} \mathbb{P}_{T}( X \geq 2f| m+1)^{2n+2}
    \end{split}
  \end{equation}
\end{proposition}
The estimate for the expected value provides a fast computation of the expected value without the need of computing many binomial-coefficients which is in general slow if $n$ gets big.
\subsection{TCP vs. UDP}
In the following we derive the probability $\mathbb{P}_{T}$ for UDP. Assume that the probability $p(l)$ of encountering a packet loss when a message (with length $l$) is transmitted using UDP is given. Then $p_{l,UDP} := p(l)$ because UDP does not bother whether a message has been successfully sent. The probability of receiving $j$ out of $n$ messages using UDP reads as
\begin{equation}
\label{eq:udp:k-out-of-n-messages}
\mathbb{P}_{UDP}(X = j| n) =  \left(
    \begin{array}{c}
      n \\
      j
    \end{array}
  \right) p(l)^j (1-p(l))^{n-j}.
\end{equation}
For UDP we use $\mathbb{P}_{UDP}$ as an instantiation of $\mathbb{P}_{T}$ in Equation~\ref{eq:probability_nodes}.
The service provider may use the expected value $\mathbb{E}[S, J \geq 2f+1, K \geq 2f+1, M \geq 2f+1]$ to decide whether it should switch from a UDP based transmission to TCP. A criteria for switching the transport protocol could be $\mathbb{E}[S, J \geq 2f+1, K \geq 2f+1, M \geq 2f+1] < 2f+1$  or Equation~\ref{eq:expected-value} because at least $f+1$ (in the best case) or $2f+1$ (in the worst case) replies are needed by the client accepting the transaction. With TCP we gain reliable connections at the expense of (even more) delay (and time until a phase completes). Thus, we want to minimize the impact of re-transmissions. Therefore, we would like to know how many re-transmissions of a single message do we need on average such that $\mathbb{E}[S, J \geq 2f+1, K \geq 2f+1, M \geq 2f+1] \geq 2f+1$.
Using TCP we assume a constant transaction success probability of one, assuming an infinite number of re-transmissions, but employing higher latency because of the acknowledgment mechanism and potential re-transmissions.

In order to shed some light on the probability and expected value of TCP re-transmissions, we assume that TCP connections are already set up and we only account for the transmission of data segments/messages. Again, we assume that the probability of successfully transmitting a packet of length $l$ over the wire/channel is given by $p(l)$.
However, a segment is only successfully transmitted using TCP if we receive an acknowledgment (ACK) otherwise a time-out will trigger, and a re-transmission of the segment will be initiated. Therefore, both (segment + ACK) have to be transmitted successfully. We do not consider any extensions of TCP. A message may be divided into several segments which all have to be successfully transmitted. The probability of successfully transmitting a segment reads as
\begin{equation}
\begin{split}
\label{eq:tcp:transmission:segment}
\nonumber
&\mathbb{P}( \neg M | p) = p(l)(1-p(ACK)) + (1-p(l))\\
&\mathbb{P}( M | p) = p(l)p(ACK).
\end{split}
\end{equation}
If $p(l) \approx p(ACK)$ then we have $\mathbb{P}( M | p) = p(l)^2$, in general we have $p(l) \leq p(ACK)$ and we obtain $\mathbb{P}( M | p) \geq p(l)^2$.
We derive the probability of successfully transmitting a segment with a certain number of allowed re-transmissions $m \in \mathbb{N}_0$ by Equation~\ref{eq:tcp:retransmission:segment}.
In order to derive the probability of successful transmitting a TCP segment, we model this process $(X_n)_{n \in \mathbb{N}}$ by a Markov chain with the state space $\Omega=\{1,2\}$ with the following transition matrix
\begin{equation}
\nonumber
P = \begin{pmatrix}
1 & 0 \\
\mathbb{P}( M | p) & 1-\mathbb{P}( M | p)
\end{pmatrix}.
\end{equation}
According to the Kolmogorov -- Chapman equation we obtain for $\mathbb{P}_{RETCP}(M | m,p)$
\begin{equation}
\label{eq:tcp:retransmission:segment}
\mathbb{P}(X_m = 1 | X_0 = 2) = P_{2,1}^m = \mathbb{P}( M | p) \sum_{k = 0}^{m} (1-\mathbb{P}( M | p))^k.
\end{equation}
Equation~\ref{eq:tcp:retransmission:segment} can be easily verified by applying induction.
%
\begin{proposition}
Let $(\Omega, \mathcal{A}, \mathbb{P}_{RETCP})$ be a probability space, where $\Omega = \{M, \neg M\}$, with the states accounting for a successful and not successful transmission of a TCP segment. Where, $(\mathbb{P}_{RETCP})$ is conditional probability measure given a certain number of re-transmissions $m \in \mathbb{N}_0$. Then the following holds
\begin{equation}
\nonumber
\lim_{m\rightarrow \infty} \mathbb{P}_{RETCP}(M | m,p)  = 1.
\end{equation}
\end{proposition}
%
%
\begin{corollary}
Equation~\ref{eq:tcp:retransmission:segment} can also be written as
\begin{equation}
\nonumber
\mathbb{P}_{RETCP}(M | m)  = 1 - (1-\mathbb{P}( M | p))^{m+1}.
\end{equation}
\end{corollary}
A message sent by our BFT solution may be split up into several TCP segments. Assuming an \textit{i.i.d.} packet loss, the success probability of a message which is divided into $u$ different segments finally reads as
\begin{equation}
\label{eq:tcp:transmission:message}
\begin{split}
p_{l,TCP} := \mathbb{P}\left( \bigcap_{j=1}^{u} M_j | m,p \right) = \prod_{j=1}^{u} \mathbb{P}_{RETCP}(M_j | m,p) = \\
(1 - (1-\mathbb{P}( M_1 | p))^{m+1})^{u-1} (1 - (1-\mathbb{P}( M_k | p))^{m+1}),
\end{split}
\end{equation}
there are $k$ segments where $k-1$ are of the same size and the $k$-th segment may have a smaller length than its predecessors. Then the probability that a replica receives $k$ messages using TCP reads as
\begin{equation}
\nonumber
\begin{split}
\mathbb{P}_{TCP}(X = k| n, p_l) = \left(
    \begin{array}{c}
      n \\
      k
    \end{array}
  \right)  p_{l,TCP}^k (1-p_{l,TCP})^{n-k}.
\end{split}
\end{equation}
The probability that $i$ replicas receive at least $2f$ messages (excluding the self-message) reads as
\begin{equation}
\label{eq:tcp:consens}
\nonumber
\begin{split}
&\mathbb{P}(Y=i, X \geq 2f)  =  \\
&\left(
    \begin{array}{c}
      n \\
      i
    \end{array}
  \right) \mathbb{P}_{TCP}(X \geq 2f | n-1)^i (1-\mathbb{P}_{TCP}(X \geq 2f| n-1))^{n-i}.
\end{split}
\end{equation}
\begin{proposition}
\label{proposition:tcp:retransmissions}
Let $1 \leq f \leq \lfloor \frac{n-1}{3} \rfloor$, then we obtain the following inequality and lower bound on the needed re-transmissions:
\begin{equation}
\begin{split}
\mathbb{E}_{TCP}[S, J \geq 2f+1, K \geq 2f+1, M \geq 2f+1] \geq \\
n(1-(1-p(l)^2)^{r+1})^{u \cdot n+(2n-2)\cdot (n-1)}
\end{split}
\end{equation}
In order to have at least $2f+1$ replicas that successfully reply to the client we need at most
\begin{equation}
\label{eq:tcp:retransmissions:number}
\begin{split}
r = \left\lceil \log_{1-p(l)^2}\left(1-\left(\frac{2f+1}{n}\right)^{(u \cdot n+(2n-2)\cdot (n-1))^{-1}}\right) - 1 \right\rceil
\end{split}
\end{equation}
re-transmissions using TCP.
\end{proposition}
Proposition~\ref{proposition:tcp:retransmissions} provides a rule of thumb for the number of needed re-transmissions for each TCP transmission such that in the end the client receives enough replies. We may also use the insights gained by Equation~\ref{eq:tcp:retransmissions:number} for UDP. If we set $\mathbb{P}( M | p) = p(l)$ we have the case of UDP. In this case we have an estimate on how often each BFT node hast to duplicate (incl. sending) a message. Thus, before switching to TCP, the BFT system may try to send each message $r$ times.

\subsection{Exploring the design space}

In the following we discuss the most important parameters and improvements to tune system deployment to optimize the performance.

\subsubsection{Forward Error Correction (Replication Code)}

To improve the probability for a packet being transmitted successfully without the introduction of handshake protocols like TCP we could apply forward error correction (FEC) mechanisms.
The simplest way would be to apply, replication codes, which send the data multiple times.
In case of immediate re-transmission with UDP a new $p_{l_2}$ and $p_{l_3}$ for having an additional re-transmission (replication code) or two additional immediate re-transmissions respectively would decrease the packet loss substantially for our channel model with i.i.d. loss ($p_{l_2} = 1 - (1-p_l)^2 = 2 p_l - p_l^2$ and $p_{l_3} = 1 - (1-p_l)^3$)


\subsubsection{Additional Redundancy in Nodes}

An alternative solution would be the use of additional nodes beyond the optimal $3f+1$ robustness bound.
For the standard case with reliable channels it does note make sense to go beyond the optimal number of nodes, because no robustness is gained.
However, from a performance perspective, increasing the amount of nodes $3f+1+x$ leads to higher success probabilities in the UDP case and could improve system performance if switching to TCP could be pushed to higher error rates or even avoided for the expected communication channels.


%


%% file: conference/sections/eval.tex
\section{Performance Evaluation}
\label{sec:performance_evaluation}
In order to investigate the performance of the proposed approach and to validate the theoretical results we simulated the BFT protocol as described in Section~\ref{sec:model}.
We selected OMNet++ 5.6 as the underlying simulation environment~\cite{OMNET} and use INET 3 as the network simulator~\cite{INET} on top of which we implemented the altered PBFT protocol using TCP and/or UDP as transport protocol for exchanging messages on the application layer.
We use a simplified topology where $n$ replicas are connected through a router.
Additionally, we benchmarked a real PBFT implementation developed in a project for multi-cloud storage to verify the results from the event simulation and test improvements.

\subsection{Model Validation}

\begin{figure}[tbh!]
\centering
\includegraphics[width=0.48\textwidth]{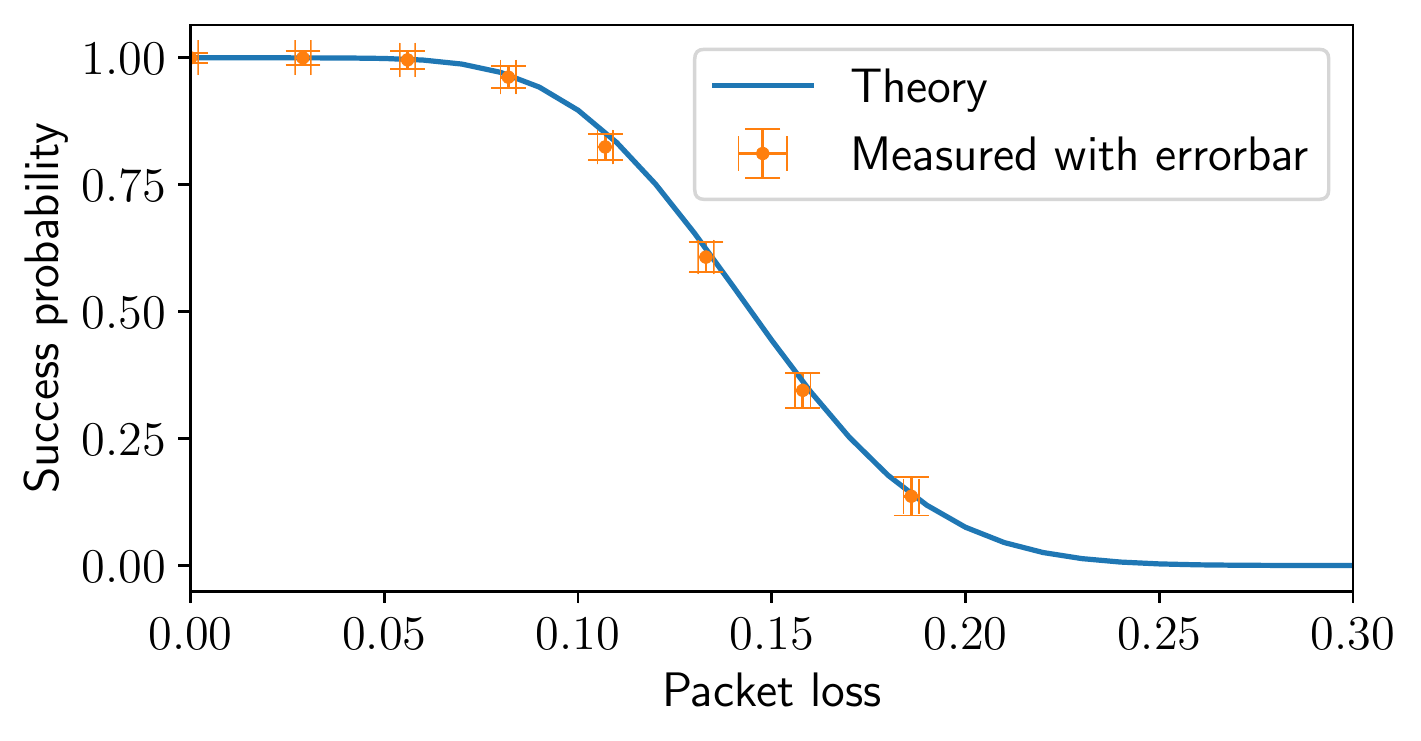}
\caption{Transaction success probability as a function of packet loss obtained by experiments vs. Equation~\ref{eq:probability_nodes} using UDP.}
\label{fig:udponlyvstheory}
\end{figure}

For the first experiment we set the bandwidth of each link (between node and router) to 100 Mbps, and the delay is truncated normal distributed (always $\geq 0$) with mean 20ms and a variance of 5ms. We varied the bit error rate of the channel from 0 to $13\cdot 10^{-5}$ in $10^{-5}$ steps and measured the actual packet loss seen at the transport layer. We used 20 replicas, a message size of 128 bytes, and we assumed the maximum number of faulty nodes (6 in the case of 20 nodes). For each simulation run we did 100 requests and for each simulation parameter configuration we did 20 repetitions. Figure~\ref{fig:udponlyvstheory} depicts the probability using the model provided in Equation~\ref{eq:probability_nodes} ($P_{succ}:=\mathbb{P}(S \geq 2f+1, J\geq 2f+1, K \geq 2f+1, M \geq 2f+1)$) and the data obtained by the experiment. It is evident that the theoretical model fits the observed experimental data.

\begin{figure}[tbh!]
\centering
\includegraphics[width=0.48\textwidth]{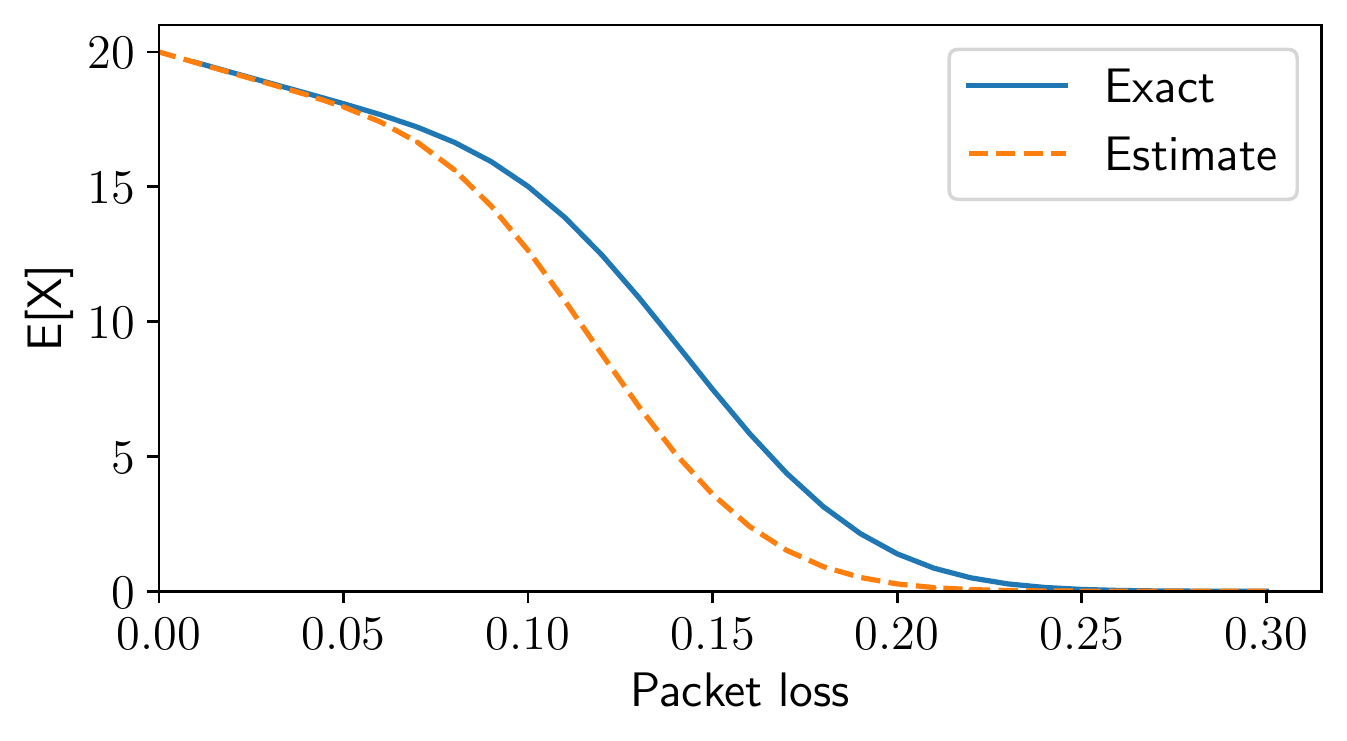}
\caption{Comparison between the exact expected value (black line) of replicas replying to the client being verified by experiments and the estimate given in Equation~\ref{proposition:expected_value} (dashed red line) for the case of UDP transmission. The parameters in order to obtain these expected values are the same as for the experiment.}
\label{fig:upperlowerbounds}
\end{figure}

Even if the theoretical model fits the experimental data it is not feasible to work with the exact formula for larger deployments, especially if we want to know how many nodes are at least expected to reply to the client. Figure~\ref{fig:upperlowerbounds} provides a graphical comparison between the exact result and the estimate given in Equation~\ref{proposition:expected_value} and shows a good fit between model and simulation.
%

The relation is mainly governed by the length of the packets transmitted.
The length of the packets are rather short, however, to cope for possible different
packet lengths we use the packet error rate for comparison which makes the results independent of variations in packet length.


\subsection{Simulation Results}

To better understand and improve the UDP behavior we explore the design space available to improve success rates and analyze their impact on the latency.
Two immediate and easy to realize options exist for the improvement of the success probability of individual transactions $P_{succ}$.
One is to increase the redundancy of nodes and the other to better cope for channel losses by means of forward error correction (FEC).

\begin{figure}[t]
\centering
\includegraphics[width=0.48\textwidth]{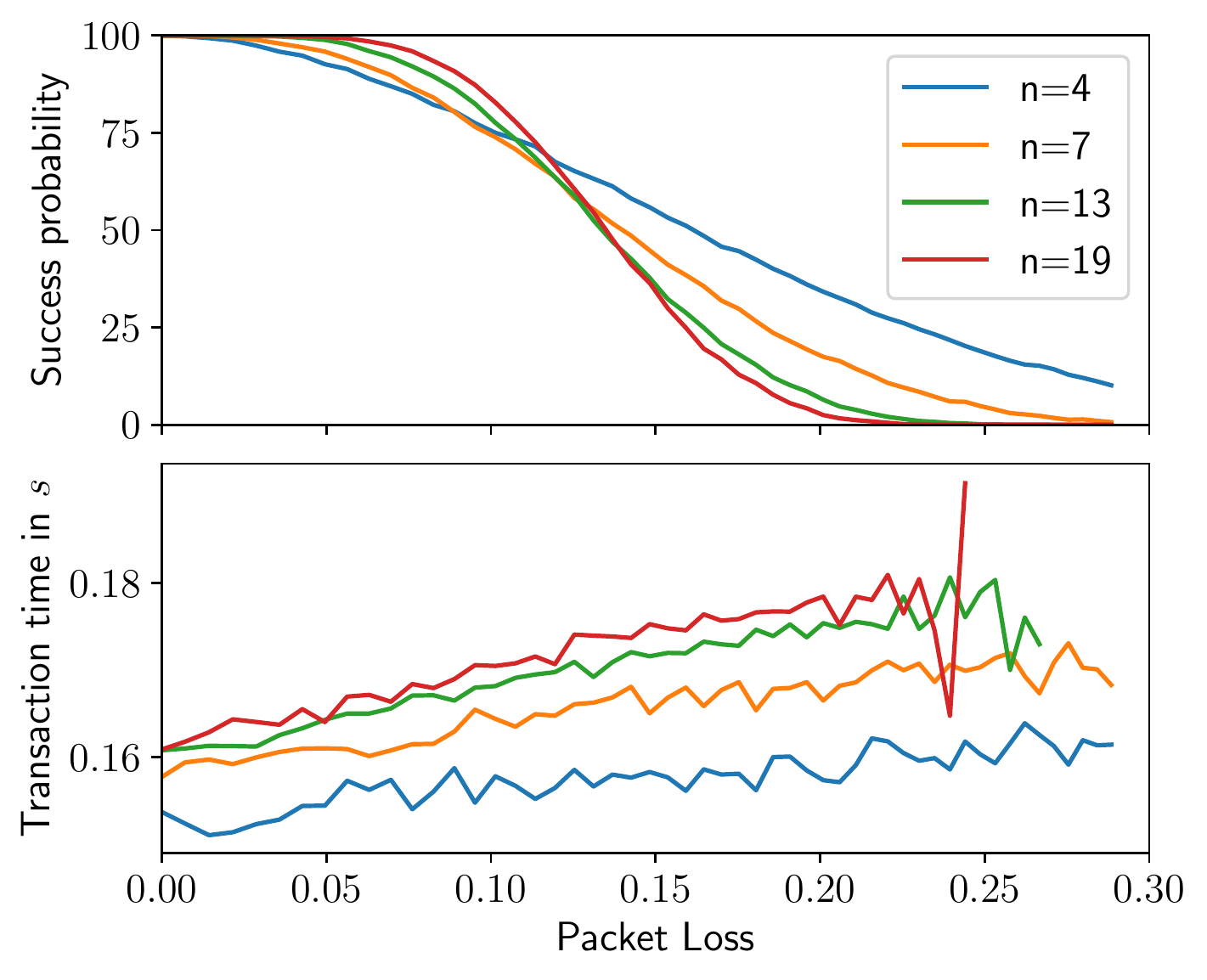}
\caption{Success probability over increasing packet loss for UDP with different f and minimum node configuration $n=3f+1$.}%
\label{fig:udpnoderedundancyf0}
\end{figure}

\begin{figure}[t]
\centering
\includegraphics[width=0.48\textwidth]{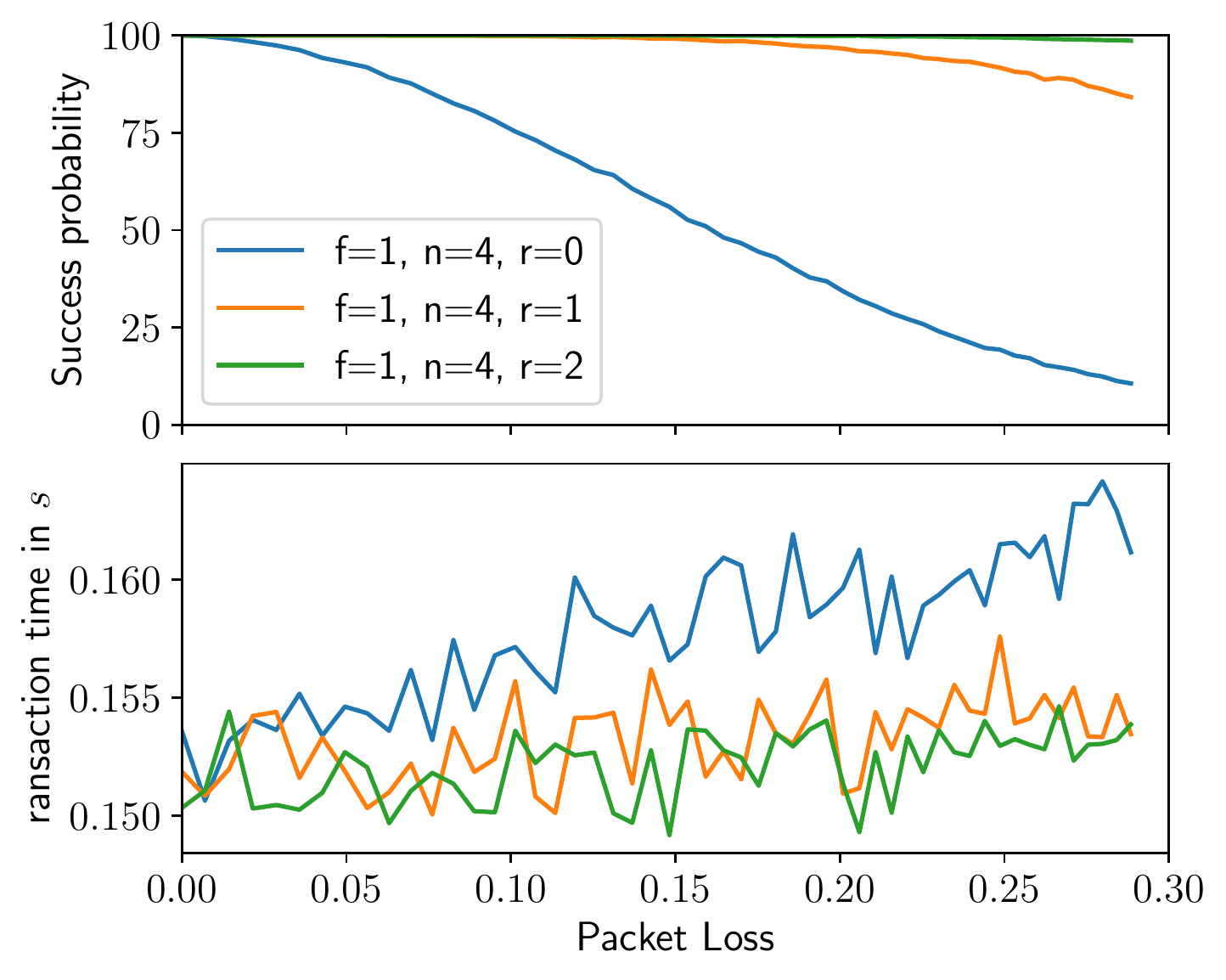}
\caption{Success probability over increasing packet loss for UDP with $f=1$ and increasing repetitions.}%
\label{fig:udprepetitionsf1}
\end{figure}

To prevent transactions from failing by losing synchronization at certain nodes, increasing the number of nodes seems a good way to increase resilience.
However, the main configuration parameters of a BFT system ($n$, $f$) cannot be freely chosen and have to fulfill certain requirements.
In general, a setting with $n=3f+1$ is believed to be optimal and typically used, as the quorum size is also minimal with $2f+1$.
We therefore compared settings with different robustness $f$ from a performance point of view and for the suitability of UDP.
The results are shown in Figure~\ref{fig:udprepetitionsf1}, and it can be seen that with increasing number of nodes $n$, the success probability $P_{succ}$ also increases.
For settings with an intermediate number of nodes (e.g. $n>=19$) we see high transaction success even for substantial packet loss, which indicates that application of UDP is practical.
Furthermore, as expected the transaction times are much better with UDP compared to protocols using acknowledgements and only slightly increases with higher packet loss and number of nodes.

If FEC is used, repetition codes are the most efficient solution in our case, as the amount of packets should be kept low and only short messages are exchanged in multiple rounds.
The effect of repetition codes is shown in Figure~\ref{fig:udprepetitionsf1}.
As expected it raises $P_{succ}$ substantially by reducing the effective packet loss on the channels through proactive retransmission of packages.
This comes at the cost of an (unnecessary) increase of messages transmitted.
Interestingly, the overall transaction time is not affected if enough bandwidth is available and the good timing behavior is maintained in all situations.

Given an accurate channel model and some bandwidth left on the network, this method turned out to be the most effective.
However, if the channel changes behavior or is not known at all, this approach could lead to completely different results, e.g., for burst failures this FEC strategy would fail.
Additionally, overhead on the network is produced and it should only be used if enough bandwidth is available and no additional congestion is induced.

\begin{figure}[t]
\centering
\includegraphics[width=0.48\textwidth]{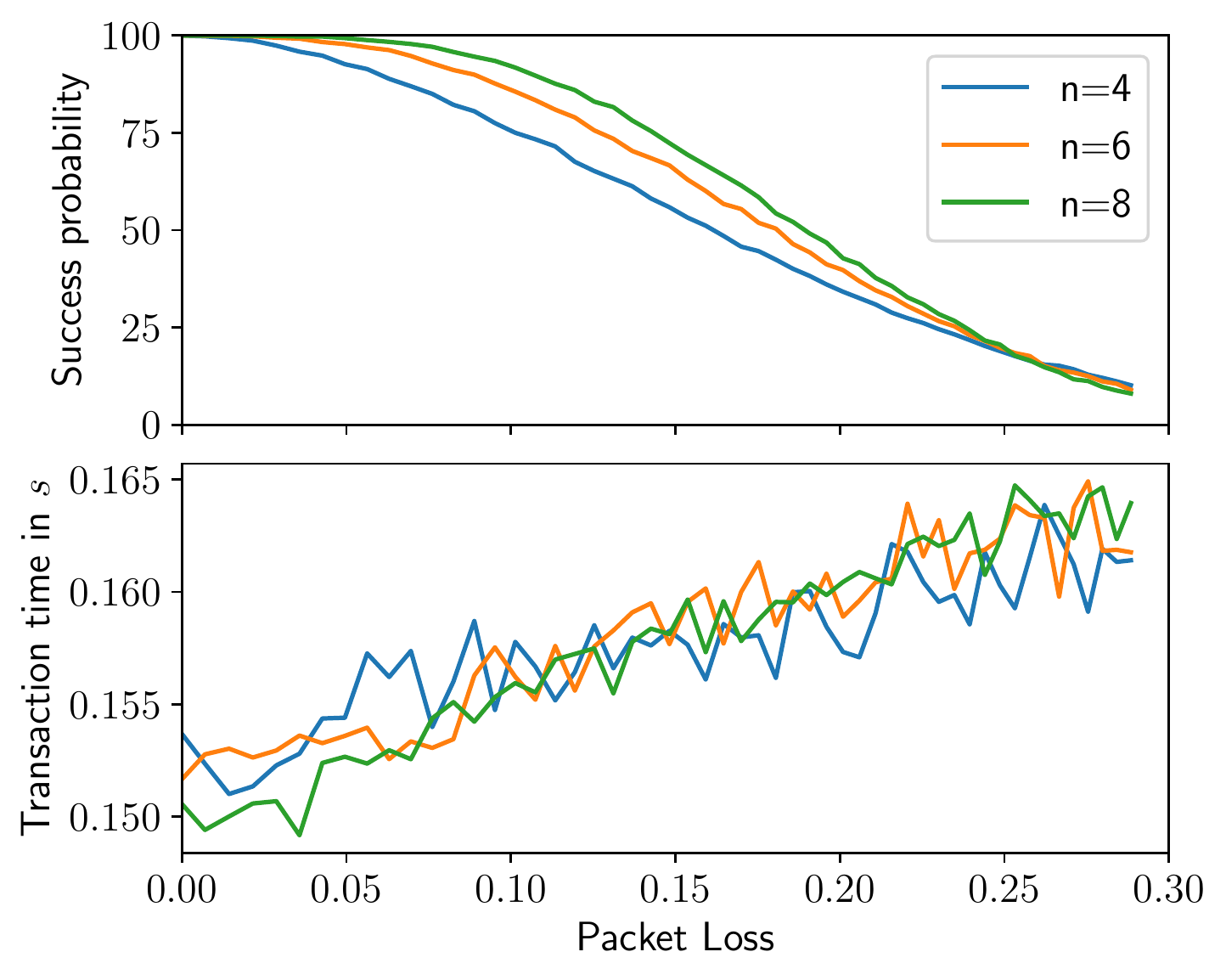}
\caption{Success probability over increasing packet loss for UDP with $f=1$ and increasing node redundancy}%
\label{fig:udpnoderedundancyf1}
\end{figure}

Finally, besides the evident options presented above, it is natural to ask if going beyond optimal configurations of $n=3f+1$ could make sense from a performance point of view, although not necessary from a robustness perspective.
We suspected that adding additional nodes could help to improve UDP usage even with certain packet loss, but is was not clear how it would impact the overall latency and how big the improvement in success probability would be.
In Figure~\ref{fig:udpnoderedundancyf1} we show the results of this analysis.
With additional nodes the success probability with lossy links can be increased and at the same time we get even shorter transaction times.
The effect is best seen for small configurations which can benefit from this idea.
Nevertheless, because PBFT is a quorum based protocol, nodes have to be added pairwise.
Adding a single node to an optimal configuration degrades performance, because the Byzantine quorum also increases, i.e., if more than $(n+f)/2$ servers have to be in the same phase, the servers have to wait for more PREPARE and COMMIT messages.

\begin{figure}[t]
\centering
\includegraphics[width=0.48\textwidth]{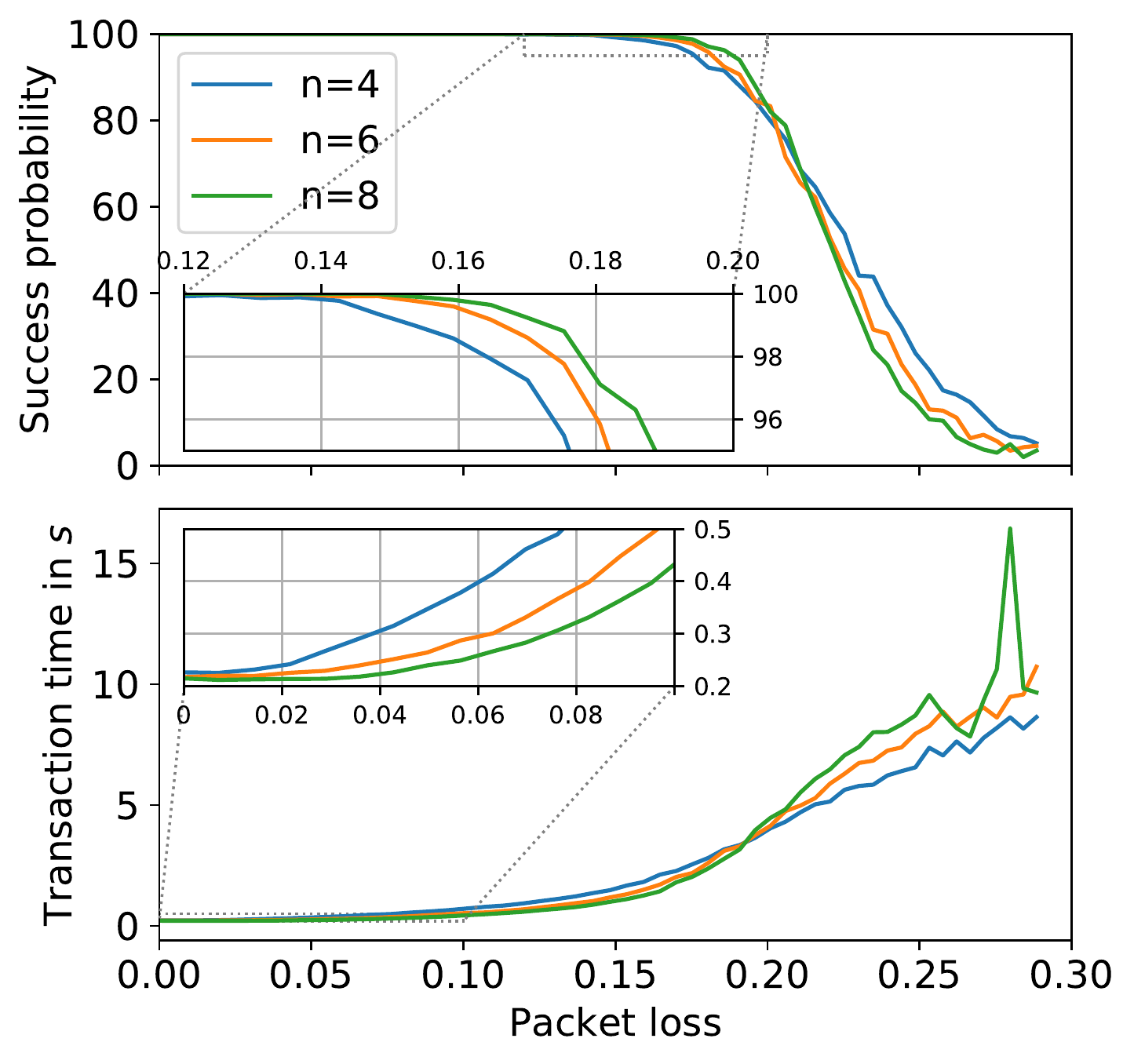}
\caption{Success probability over increasing packet loss for TCP with $f=1$ and increasing node redundancy}%
\label{fig:tcpnoderedundancyf1}
\end{figure}

Finally, in our simulations we also verified that TCP behaves worse for increasing packet loss as is shown in Figure~\ref{fig:tcpnoderedundancyf1}.
Even for no losses the transaction time was already almost twice as high as with UDP.
This can be easily explained by the basic nature of TCP using acknowledgements.
Even worse, with increasing packet loss the transaction time started to rise to unexpectedly high values in the seconds range and due to timeout behavior we even saw some transactions not finishing.
This result confirmed our findings from the first implementations mentioned in Section~\ref{sec:motivation}.

Although TCP is an extremely versatile and attractive protocol for many situations to build reliable channels over unreliable ones, for the BFT type of interactive protocols with many short messages sent among nodes it turned out to be not a good fit.
This is also aligned with our intuition of TCP being throughput optimized for channels with high bandwidth-delay product.
Nevertheless, in situations with a lot of uncertainty about the channel and high losses it can be a valuable tool to increase the transaction rate in such rough conditions.
Surprisingly we also found that the success probability was not 1 in all situations, and even with long timeouts some of the transactions did not complete in scenarios with higher packet loss.
This is because of the limit of 12 retransmissions in the TCP implementation of INET.

Finally, we also tried to compare different TCP types to show their behavior, but we could find no significant differences between the algorithms implemented in INET (Tahoe, Reno, New Reno).
This may be due to a known problem of this framework~\cite{issue}.

\subsection{System Measurements}

In addition to the simulation, we also performed measurements on a real implementation developed in a previous project.
To establish similar conditions for our comparison we opted for a local single Linux PC deployment where each node was run as a separate instance and the local network stack was used for communication.
To evaluate different networking conditions the Linux \textit{netem} kernel module~\cite{Hemminger2005} was used to provoke packet delay and network loss.
This setup provided the stable and controllable environment we needed to verify the results of the simulation and the analytical model.
For the measurements the same channel settings were used as in the simulation, i.e. normally distributed network latency with 40ms mean and 10ms variance (equals 20ms mean and 5ms variance in the star topology used in the simulation) with an additional packet loss varying from 0 to 30\%.

\begin{figure}[tbh!]
\centering
\includegraphics[width=0.48\textwidth]{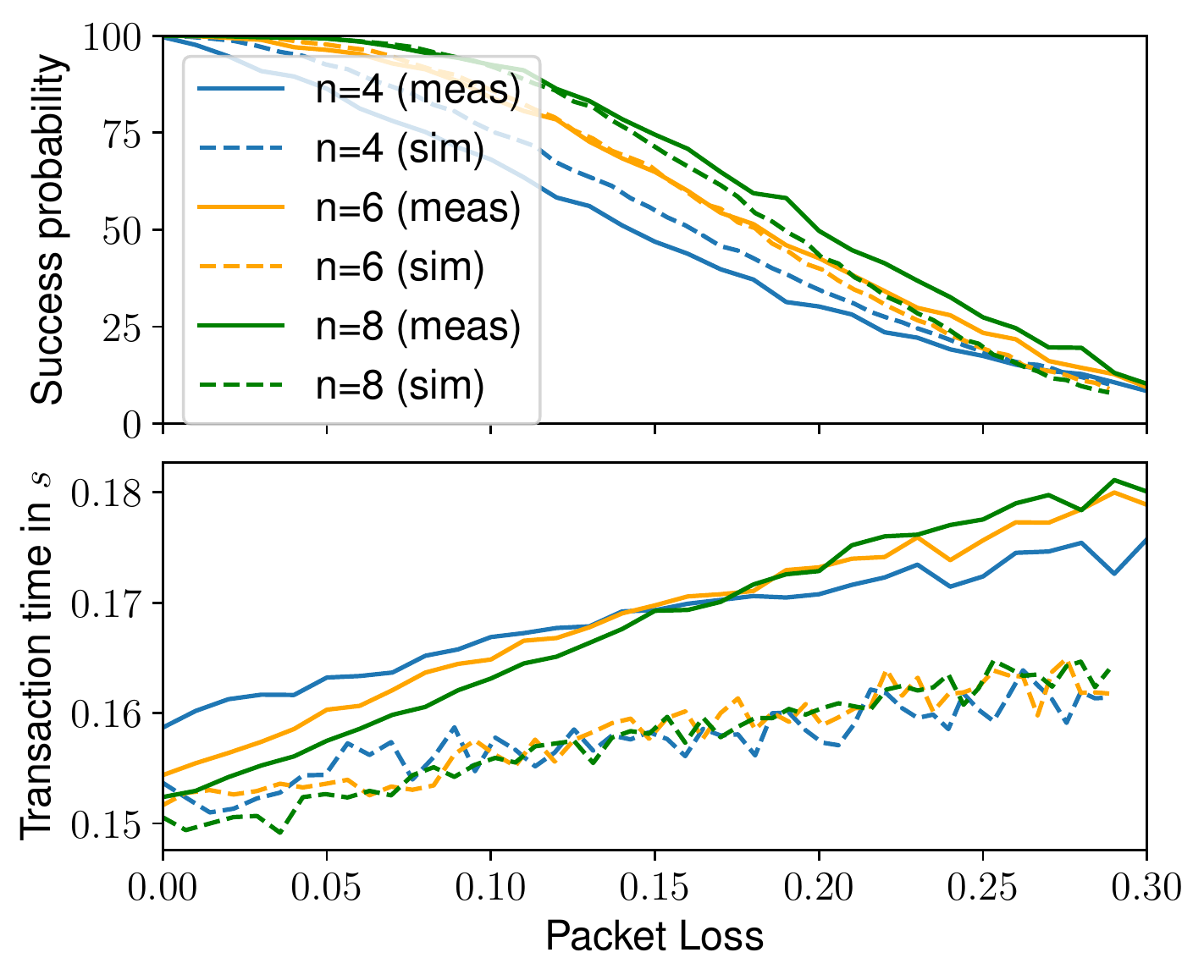}
\caption{Comparison of measured value from implementation to simulated vales for UDP. Measured values are drawn with continuous lines and simulated values dashed.}
\label{fig:udpmeas}
\end{figure}

The comparison of the measurements and the simulation is shown in Figure~\ref{fig:udpmeas}.
Overall, the measurements taken from the PBFT implementation show a very good match to the simulated results and show that model and simulation are correct and can be used to estimate performance.
The success probability in particular resembles the simulated values well.
The measured latency shows a smoother behavior over increasing packet loss corresponding to smaller variances in the measurements which can be attributed to buffering effects in the software and OS stack used.
We also found a slightly higher transaction time in the real implementation for increased packet loss, however, even for very high packet loss it was within 10$\%$ margins.

Additionally, in our protocol analysis we found that especially the PRE-PREPARE phase is susceptible to packet loss and could greatly impact the overall performance in terms of successful transaction termination.
This is due to the leader-based structure of the core view-consensus protocol in PBFT.
In such a protocol one node initializes the transactions by distributing relevant data to all other nodes, the backups.
In this phase the protocol has less redundancy compared to later phases.
Interestingly, adding redundancy by message repetition only in this phase gives a high increase in success probability with relatively low additional communication cost.
With one re-transmission in the PRE-PREPARE phase only $n-1$ packets are added, compared to $n^2$ packets per retransmission in the other phases, but the success probability can be substantially increased.
To verify this behavior we measured the increase in success probability for one and two retransmissions in the PRE-PREPARE phase.

\begin{figure}[tbh!]
\centering
\includegraphics[width=0.48\textwidth]{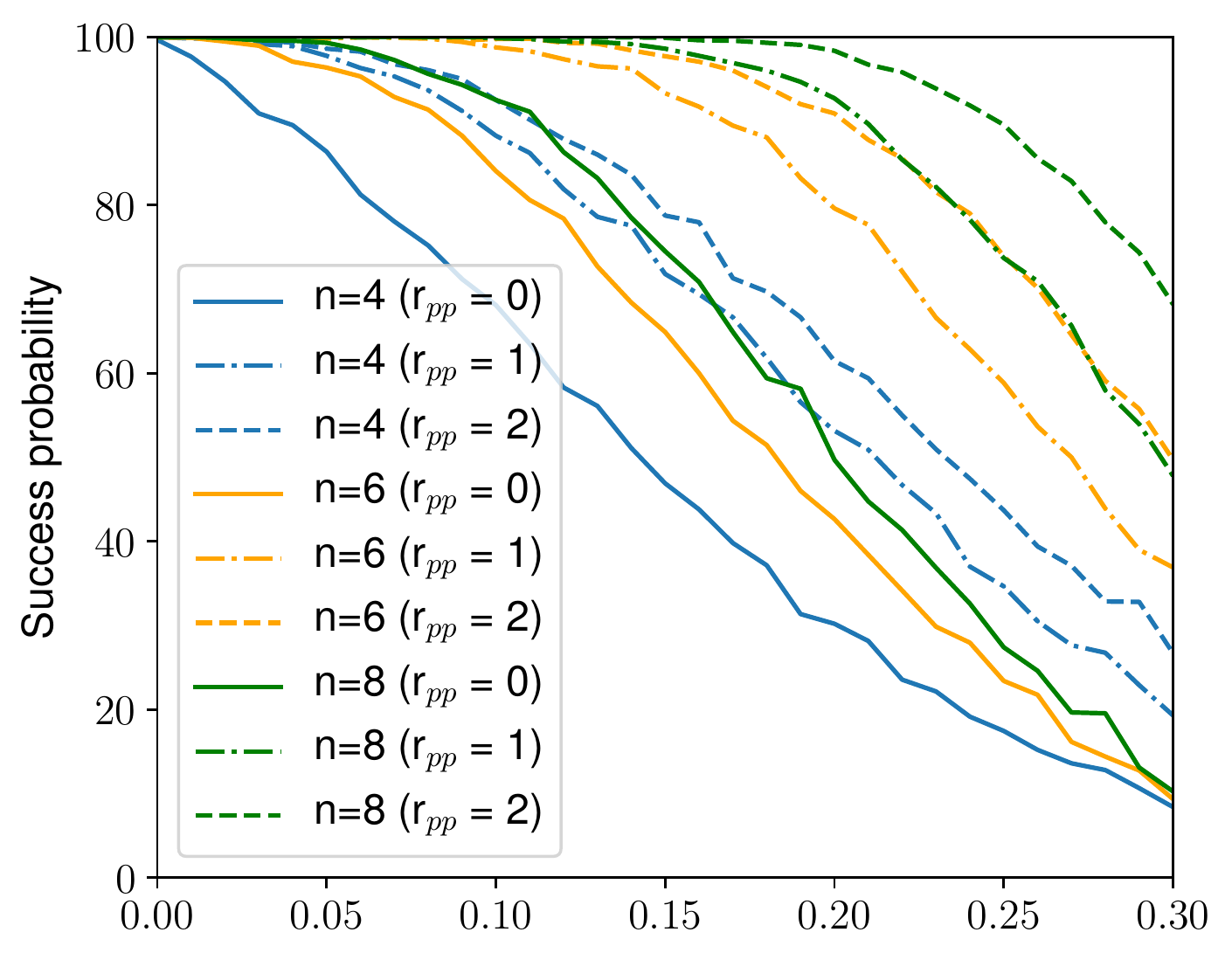}
\caption{Measured success probability with retransmission only in the pre-prepare phase. Lines without retransmission are depicted as continuous lines and results with 1 (2) retransmission of pre-prepare messages are drawn with dash-dot (dashed) style.}
\label{fig:udpmeaspreprepare}
\end{figure}

The results are presented in Figure~\ref{fig:udpmeaspreprepare}, and the data show that adding one retransmission in the PRE-PREPARE phase leads to the same or even higher $P_{success}$ as adding a full additional node for redundancy, but saves a lot of communication overhead.
Given a total of $(r_{pp}+1)n+2n^2+f+1$ messages sent in the view-consensus protocol with its three phases, with $r_{pp}$ being the number of retransmission in the pre-prepare phase, the overhead introduced with one additional retransmission is low.
For systems which tolerate one faulty node out of 4 nodes we get about $11\%$ of message overhead, with 5 nodes we see $9\%$ overhead and about $7.7\%$ overhead are required for 6 nodes.
This leads to a significant improvement compared to the communication overhead introduced by adding an additional node without retransmission to increase $P_{succ}$, i.e., a total of 53\% more messages must be sent if n is increased from 4 to 5.
Nevertheless, both measures can be combined to get UDP performance up to 5\% packet loss and more if two additional nodes are combined with retransmission in the pre-prepare phase as an example.

From this result, we see that careful design on the network layer is essential for PBFT and protocols with similar communication patterns to achieve best performance in challenging network settings.
Especially multi-cloud configurations fall in this category, but single cloud deployments with a certain level of geo-separation could also introduce substantial latencies.
As can be seen from the measurements taken at CloudPing~\cite{Matt}, latencies between continents are crucial, for example between Europe and North Americahey, where they range from $100-150\text{ms}$ (50th percentile).
Even within a single continent they are the dominating factor for BFT performance, e.g., they go up to $40ms$ (50th percentile) for servers within Europe.
Thus even intra-region BFT will face substantial latencies and has to rely on UDP for performance reasons.
However, if UDP is used, its performance should not degrade if higher packet loss is encountered and switching to TCP should be avoided if high transaction rates are required.

\subsection{Interpretation}

In general, it is desirable to use UDP and to avoid TCP wherever possible, because it leads to unacceptable performance degradation for higher error rates on the transmission channel.
Although from a robustness point of view there is no reason to use more than $3f+1$~nodes to run a PBFT system, when it comes to unreliable communication it turns out that adding nodes is a means to improve the redundancy on the network layer.
Additionally, the use of repetition codes can also lead to significant performance improvements as UDP can be used over TCP even in situations with increased packet loss.
If the channel behavior is known in advance we recommend to configure the deployment adequately to stay in the UDP regime.
In the end, for our type of application a dedicated network protocol would be desirable which adaptively optimizes retransmissions and other parameters without increasing latency.

\subsubsection{Adaptive and hybrid network layer}

From the structure of the communication pattern it turned out that unreliable channels have different impact in different phases.
A node missing a single PRE-PREPARE message could already be out of sync for the current transaction, contrary if $f$ PREPARE messages do not arrive, it will still have enough information to proceed.
This shows that especially the first broadcast from the primary is relatively more important than the rest of the messages and measures taken to increase its probability of success will have a disproportionate impact on the success of the whole transaction.
It could therefore make sense to use TCP only for this phase, or, as we have done, to pro-actively repeat this message once or twice.

\subsubsection{Byzantine case}

If $f$ nodes really behave fully malicious, their messages are ignored by the honest nodes if they do not follow the protocol.
Therefore, the best they can do to slow down transactions---and therefore slow down service time---is to delay their transmissions or remain silent.
For the network layer this would mean that no redundancy is left to cope with packet loss as all $2f+1$ honest nodes have to reach the final state for the transaction to complete and in this case packet loss would be fatal.
However, by increasing the redundancy beyond $3f+1$ nodes we reach the same regimes as presented above.
In fact if $5f+1$ nodes are used we reach in the worst case similar success probabilities, because such a system would require a $3f+1$ quorum and leave $2f$ overall redundancy in the system, i.e. $f$ Byzantine nodes and $f$ honest nodes whoes message do not need to arrive.
However, this is only true if the adversary does not have access to the channels between honest nodes, which was the assumption we started from.
Alternatively, the implementation can always fall back to TCP and therefore emulate reliable channels over unreliable ones, if the packet loss or the number of node failures is too big for UDP usage.
In essence, the safety property of the system is never compromised, only performance is improved in rather optimistic scenarios.

%

%
%

%% file: conference/sections/conclusio.tex
\section{Conclusions and Future Work}
\label{sec:con}

In this work we present the impact of packet loss and latency as well as transport protocols on the performance of BFT systems.
We provide an analytical framework and validate three obtained analytical formulas by simulations.
We further explored the design space available for PBFT deployments to optimize performance and the results have also been compared to a real implementation.
However, we did not yet complete our discussion where we would like to pose questions on the transaction time if we employ reliable and/or unreliable network communication.
We also considered only basic transactions and did not incorporate view-change protocols and garbage collection mechanisms.
For a complete picture of the overall performance these steps should be also analyzed and optimized.
Thus, we have to leave this investigation to future work.
Additionally, it is worth studying variants of PBFT, and related distributed protocols in general, that use slightly modified communication patterns but could benefit from our treatment.